\documentclass[11pt, twoside]{article}                        

\usepackage{graphicx}                   
\usepackage{asp2004}                     

\begin{document}

\title{A Short Characteristic Solution for the 2.5D Transfer Equation in
         the Envelopes of O and B Stars}

\author{J. Zsarg\'{o} and D. J. Hillier}
\affil{ Dept. of Physics and Astronomy,
University of Pittsburgh, 3941 O'Hara Str.,
Pittsburgh, PA 15260, USA}

\author{L. N. Georgiev}
\affil{Instituto de Astronomia,
Universidad Nacional Autonoma de Mexico (UNAM),
Apartado Postal 70-264, Mexico 04510, D. F. MX}

\begin{abstract}
We discuss work toward developing a 2.5D non-LTE radiative transfer code.
Our code uses the short characteristic method with modifications to
handle realistic 3D wind velocities.
We also  designed this code for parallel computing facilities by representing
the characteristics with an impact parameter vector {\bf p}.
This makes the interpolation in the radiation angles unnecessary and allows for an
independent calculation for characterisitcs represented by different {\bf p} vectors.
The effects of the velocity field are allowed for by increasing, as needed, the
number of grid points along a short characteristic.
This allows us to accurately map the variation of the opacities and emissivities
as a function of frequency and spatial coordinates.
In the future we plan to use this transfer code with a 2D non-LTE stellar atmosphere
program \citep{geo04} to self-consistently solve for level populations,
the radiation field and temperature structure for stars with winds and without
spherical symmetry.
\end{abstract}

\section{Introduction}\label{section:intro}

Modeling the circumstellar envelopes of O and B stars is a complex
nonlinear problem.
The non-LTE level populations, the (magneto-) hydrodynamics,
and the radiation field are strongly coupled.
Thus, an iterative procedure is needed to achieve a consistent
solution.
An essential constituent of this procedure is the availability of an
accurate and fast radiative transfer code.

Progress in computer technology and the availability of fast
numerical methods now allow the development of such
codes for detailed study of 2D and 3D envelopes.
There are several possible avenues to follow.
The most straightforward is to solve the general radiative
transfer equation
\begin{equation}\label{eq:RT}
 \underline{\bf n}  {\bf \nabla}  I=  - \chi \left[ I -  S \right]
\end{equation}
and calculate the radiative transition rates using the solution.
In the above $\underline{\bf n}$ is the direction of the radiation, $I$
is the specific intensity, $\chi$ is the opacity, and $S$
is the source function (functional dependence is not indicated for clarity).
Alternatively, one could also use the moment equations, derived from
Eq.~\ref{eq:RT},
and solve directly for the moments \citep{hil98} which set the transition rates;
or use Monte-Carlo simulation to solve the radiative transfer equation and calculate
estimators of the transition rates \citep{luc99}.
We decided to use the first approach because of its simplicity and since it provides a
reasonable compromise between numerical efficiency and flexibility.

A simple iteration between the radiative transfer and the rate equations is
not a wise choice for the iterative procedure.
This is the so-called ``$\Lambda$-iteration'' which is  notorious for
convergence difficulties when the optical depth is large.
Convergence is ensured, however, by using the Approximate Lambda Iteration
\citep[ALI, see e.g.,][]{ryb91, hub92} which takes some coupling of the radiation
and populations into account by using an invertible Approximate Lambda
Operator (ALO).
In our 2D code we use the local contribution at every spatial point to construct
the ALO because it is easy to calculate and has acceptable convergence
characteristics.
The actual implementation of the ALI procedure into our full non-LTE
model atmosphere will be discussed in \cite{geo04}.

\section{Description of the Transfer Code}\label{section:code}

\begin{figure}
\includegraphics[totalheight= 6cm, keepaspectratio= true, angle= 270]{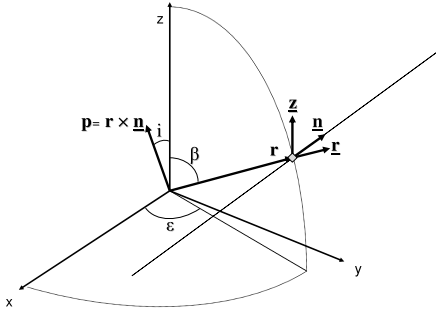}
\caption{
The definition of our fundamental coordinate system (r, $\beta$,
$\epsilon$) and the vectors that are necessary for our notations.
Unit vectors $\underline{\bf n}$,  $\underline{\bf z}$, and $\underline{\bf r}$
are describing the characteristic (long thin line), the positive z axis, and the
radial direction, respectively.
The impact parameter vector {\bf p} and $\underline{\bf n}$
are two alternative ways to define a characteristic.
}
\label{fig1}
\end{figure}

The optical depth and  the formal solution of Eq.~\ref{eq:RT} at any
position, $s$, along a ray  are
\begin{eqnarray}\label{eq:FS}
\tau_{\nu}= \int_{0}^{s} \chi ds' & ~~{\rm and}~~ &
I(\tau_{\nu})= I_{BC} \, e^{- \tau_{\nu}} \; + \;  \int_{0}^{\tau_{\nu}} S(\tau') \,
e^{\tau' - \tau_{\nu}} \, d\tau'  \; ,
\end{eqnarray}
respectively.
The intensity can be calculated by specifying I$_{BC}$ at the
up-stream end of the ray (or characterisitic)  and by evaluating two integrals.
For this purpose, we use the ``Short Characteristic'' (SC) method, first explored
by \cite{mih78} and \cite{kun88}.
This method requires the evaluation of the integrals only between the point of
interest and the closest cell boundary and uses the calculated intensities at
other grid points to interpolate $I_{BC}$.
In the spherical coordinate system the directional variation of the intensity is
normally described by the radiation coordinates $\theta$ and $\phi$,
which are defined by
\begin{eqnarray}\label{eq:radcoor}
cos(\theta)= \underline{\bf n} \cdot \underline{\bf r} &
~~~{\rm and}~~~ &
sin(\beta) \cdot sin(\theta) \cdot cos(\phi)= \left[ \underline{\bf n} \times \underline{\bf r}
\right] \cdot \left[ \underline{\bf r} \times \underline{\bf z} \right]   \; ,
\end{eqnarray}
respectively (see Fig.~\ref{fig1} for definitions).
Unfortunately, $\theta$ and $\phi$ vary along a characteristic so using the same $\theta$ and
$\phi$ grid for all spatial points would require interpolations in these angles.
To avoid this additional interpolation we describe a characteristic with
\begin{equation}
{\bf p}= {\bf r} \times \underline{\bf n} \; ,
\end{equation}
which we call the ``impact parameter vector'' (see Fig.~\ref{fig1}).
This vector describes all essential features of a characteristic and can be
considered as an analog of the orbital momentum vector in two body problems.
Its absolute value p= $|${\bf p}$|$ is the traditional impact parameter while its
orientation defines the ``orbital plane'' of the radiation (the plane that contains
the characteristic and the origin).
Following the analogy one can define an ``inclination'' angle for this plane by
\begin{equation}\label{eq:i}
p \cdot cos(i)= {\bf p} \cdot \underline{\bf z} \; .
\end{equation}
In our code we set up a universal grid in impact parameters and in inclination
angles and calculate the radiation coordinates  by
\begin{eqnarray}\label{eq:CT}
sin( \theta ) = \frac{p}{r}  &
~~~{\rm and}~~~ &
sin( \phi )= \frac{cos(i)}{sin( \beta )}
\end{eqnarray}
for each grid point (see Fig~\ref{fig1} for definitions).

We evaluate Eq.~\ref{eq:FS} in the comoving frame of the point of interest
which is the proper frame for solving the rate equations.
To ensure that the spatial and frequency variations of the opacity and source
function are mapped properly in the integrations, we add extra integration points
to the characteristics.
The number of the extra points (at least one) depends on the ratio of the line of sight
velocity difference between the endpoints and a ``maximum
allowed velocity difference'' which is a free parameter in the code.
The opacities and source terms at every comoving frequency are then
interpoleted onto the integration points by bi-linear
approximations using the four closest spatial grid points.
It is relatively easy to construct a diagonal ALO in this evaluation scheme.

With the exception of the intensity we interpolate all quantities in first order.
However, the accuracy of this approximation is insufficient in many cases;
therefore, we introduced a rudimentary multi-grid approach.
Before evaluating Eq.~\ref{eq:FS}, we calculate opacity and source function
on a dense grid by using a sophisticated 3$^{rd}$ order approximation. Then, the
transfer equation is solved on the original grid using the dense grid for opacity
and source term interpolations.

\section{Test Results} \label{section:tests}

\begin{figure}
\includegraphics[totalheight= 9cm, keepaspectratio= true, angle= 270]{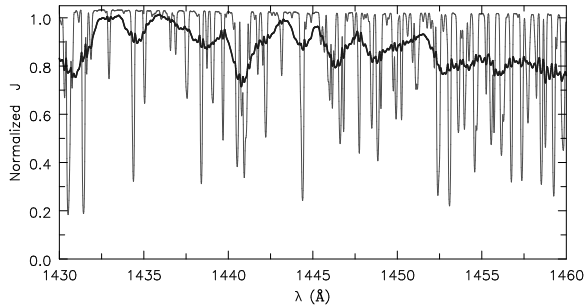}
\caption{
The J moment of the radiation field for $\beta$= 0$\deg$ (thin line) and
45$\deg$ (thick line) as a function of wavelenght (\AA ) at the outer
boundary of the envelope.
}
\label{fig2}
\end{figure}

\begin{figure}
\includegraphics[totalheight= 5cm, keepaspectratio= true,  angle= 270]{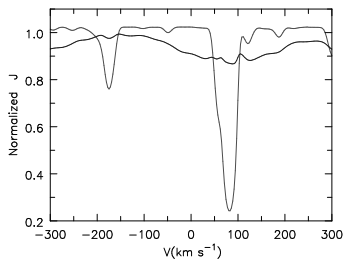}
\caption{
The spectral region around 1444 \AA~ from Fig.~\ref{fig2} as a function
of velocity (centered on 1444 \AA).
The rotational velocity in the envelope is $V(r,\beta)= 250
\cdot r_*/r \cdot sin( \beta)$ where $r_*$ is the stellar radius.
}
\label{fig3}
\end{figure}

We tested our code by reproducing spherical symmetric CMFGEN models
{\citep{hil98}}, as well as the results of an accurate 2D long characteristic
program {\citep[see][]{bus00}}.
The 2D models were static with Schuster-type inner boundary conditions and
included electron scattering iterations.
We were able to reproduce all test cases within $\sim$5\% accuracy.
In Figs.~\ref{fig2} and \ref{fig3} we demonstrate the capabilities of our code
by showing the results for a rotating stellar envelope.
The model was produced by using the opacities and emissivities from the results of a
realistic CMFGEN simulation and by introducing a rotational velocity field.
As expected the spectral lines show the rotational broadening.



\begin{thebibliography}{}

\bibitem[Busche \& Hillier(2000)]{bus00}
Busche, J. R. \& Hillier, D. J. 2000, \apj, 531, 1071

\bibitem[Georgiev {et al.}(2004)]{geo04}
Georgiev L. N., Hillier, D. J., \& Zsarg\'{o}, J. 2004, in preparation

\bibitem[Hillier \& Miller(1998)]{hil98}
Hillier, D. J. \& Miller, D. L. 1998, \apj, 496, 407

\bibitem[Hubeny(1992)]{hub92}
Hubeny, I. 1992, in The Atmospheres of Early-Type Stars, ed. U. Heber \&
C. J. Jeffery (Berlin:Springer), 377

\bibitem[Kunasz \& Auer(1988)]{kun88}
Kunasz, P. B., \& Auer, L. H. 1988, J. Quant. Spectrosc. Radiat. Transfer, 39, 67

\bibitem[Lucy(1999)]{luc99}
Lucy, L. B. 1999, \aap, 344, 282

\bibitem[Mihalas {et al.}(1978)]{mih78}
Mihalas, D., Auer, L. H., \& Mihalas, B. W. 1978, \apj, 220, 1001

\bibitem[Rybicki \& Hummer(1991)]{ryb91}
Rybicki, G. B. \& Hummer, D. G. 1991, \aap, 245,171

\end{thebibliography}
\end{document}